\documentclass[aps,prl,longbibliography,twocolumn,superscriptaddress]{revtex4-1}
\usepackage{amsmath}
\usepackage{graphicx}
\usepackage{dcolumn}
\usepackage{bm}
\usepackage[normalem]{ulem}
\usepackage{color}
\usepackage[dvipsnames]{xcolor}
\newcommand\st{\bgroup\markoverwith{\textcolor{RedOrange}{\rule[0.3ex]{2pt}{1.5pt}}}\ULon}
\usepackage{pifont}
\usepackage[geometry]{ifsym}

\usepackage{epstopdf}

\newcommand{\ii}{{\rm i}}

\newcommand{\bp}{\mathbf{p}}

\newcommand{\bff}{\mathbf{f}}

\newcommand{\bbr}{\mathbf{r}}

\newcommand{\sep}{ \ \ \ , \ \ \ }

\newcommand{\beq}{\begin{equation}}
\newcommand{\eeq}{\end{equation}}
\newcommand{\beqn}{\begin{eqnarray}}
\newcommand{\eeqn}{\end{eqnarray}}
\newcommand{\pp}{\partial}

\newcommand{\ee}{{\rm e}}

\newcommand{\fig}{Fig.\ }

\newcommand{\cO}{{\cal O}}

\newcommand{\la}{\langle}

\newcommand{\ra}{\rangle}

\newcommand{\vnab}{{\bf \nabla}}

\begin{document}

\title{Diversity of phase transitions and phase separations in active fluids}

\author{Thibault Bertrand}
\email[Electronic address: ]{t.bertrand@imperial.ac.uk}
\affiliation{Department of Mathematics, Imperial College London, South Kensington Campus, London SW7 2AZ, United Kingdom}
\author{Chiu Fan Lee}
\email[Electronic address: ]{c.lee@imperial.ac.uk}
\affiliation{Department of Bioengineering, Imperial College London, South Kensington Campus, London SW7 2AZ, United Kingdom}

\date{\today}

\begin{abstract}
Active matter is not only indispensable to our understanding of diverse biological processes, but also provides a fertile ground for discovering novel physics. Many emergent properties impossible for equilibrium systems have been demonstrated in active systems. These emergent features include motility-induced phase separation, long-ranged ordered (collective motion) phase in two dimensions, and order-disorder phase co-existences (banding and reverse-banding regimes). Here, we unify these diverse phase transitions and phase co-existences into a single formulation based on  generic hydrodynamic equations for active fluids. We also reveal a novel co-moving co-existence phase and a putative novel critical point.
\end{abstract}

\maketitle

Active matter refers to many-body systems in which each volume element can generate its own mechanical stresses \cite{ramaswamy-arcmp-2010,marchetti-rmp-2013,bechinger-rmp-2016}. As the fluctuation-dissipation relation is broken at the microscopic level, active matter can be viewed as an extreme form of far-from-equilibrium systems. Given the relevance of active matter to non-equilibrium physics and biophysics, the subject area has been rapidly expanding and many approaches have been used to study this diverse class of non-equilibrium, many-body systems. Arguably, the most generic way to investigate the emergent properties of an active matter system is to first  formulate a model based  solely on the  underlying symmetries and conservation laws of the system \cite{chaikin-book-1995}. 

This is what was done in the case of active fluids -- a class of active matter in which translation invariance holds -- in the seminal work by Toner and Tu \cite{toner-prl-1995,toner-pre-1998,toner-annphys-2005,toner-pre-2012}. Motivated by the simulation study by Vicsek et al.~\cite{vicsek-prl-1995}, Toner and Tu provided the generic equations of motion (EOM) for polar active fluids and demonstrated the existence of the polar ordered, or  collective motion, phase using a renormalization group analysis. Subsequently, a co-existence regime consisting of the ordered and disordered phases was also found, which generically separates the disordered phase and the ordered phase in typical polar active fluid models \cite{chate-pre-2008,gregoire-prl-2004,thuroff-prx-2014,solon-prl-2013,solon-pre-2015a,gopinath-pre-2012}. Numerous studies have also confirmed the Toner-Tu EOM for polar active fluids using formal coarse-graining strategies that link microscopic models of self-propelled particles and hydrodynamic level equations \cite{baskaran-pre-2008,baskaran-prl-2008,bertin-pre-2006,bertin-jphysa-2009,peshkov-prl-2012b,peshkov-epjst-2014,bertin-pre-2015}.

Concurrently, dense collections of active particles without aligning interactions were shown to spontaneously phase separate in the presence of purely steric interactions. This phenomenon is now known as {\it motility-induced phase separation} (MIPS) \cite{cates-arcmp-2015}; it was first predicted theoretically and simulationally \cite{tailleur-prl-2008,fily-prl-2012,redner-prl-2013}, and then experimentally observed \cite{palacci-science-2013,buttinoni-prl-2013}. Scalar field theories, typically based on the density field of the particles, have also been formulated to describe this process, e.g., the so-called {\it Active Model B} \cite{wittkowski-natcomms-2014,tjhung-prx-2018}.  In terms of symmetries and conservation laws, MIPS and polar active fluids are completely identical. It is therefore natural to view the emergence of the ordered phase and MIPS as properties of the same class of active systems. Through scattered efforts, recent studies have attempted to gain insight into the competition between Vicsek-like aligning interactions and steric repulsion in experiments \cite{linden-prl-2019} and in models \cite{peruani-prl-2011,farrell-prl-2012,martin-gomez-sm-2018,sese-sansa-epl-2018,caprini-prl-2020,jayaram-pre-2020} of active particles. Nevertheless, our understanding of the connections between the emergence of collective motion and phase separation is still crucially lacking. In this Letter, we elucidate the interplay between these phenomena; specifically, we unify these diverse types of phases and phase co-existences in a single formulation based on  generic hydrodynamic EOM for active fluids. In the process, we also uncover a new co-existence regime and a putative novel critical point.

\begin{figure*}
\centering
\includegraphics[width=17.2cm]{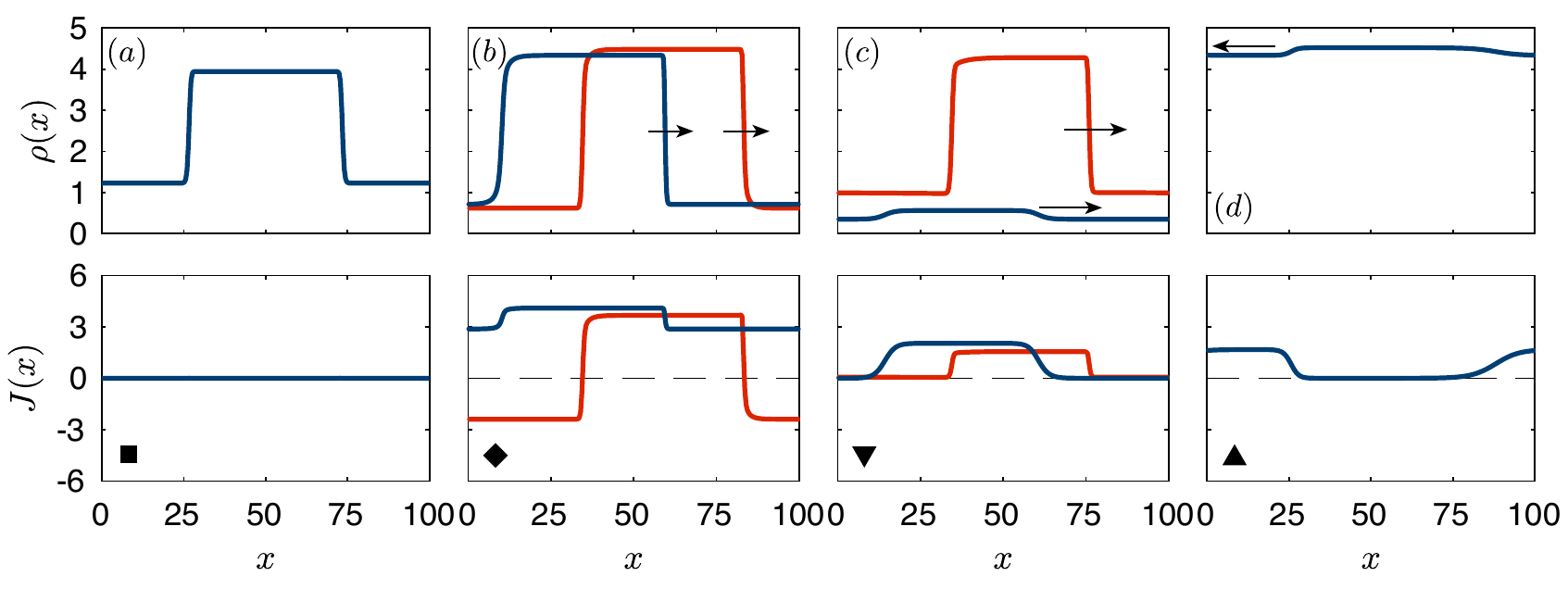}
\caption{Phase co-existence in an one-dimensional active fluid. The mass density $\rho$ (upper panels) and total flux $J$ (lower panels) profiles of the four distinct types of phase co-existences: (a) a dilute ({\bf d}) and disordered ({\bf D}) phase co-existing with a condensed ({\bf c}) and disordered ({\bf D}) phase, denoted as {\bf dD-cD}; (b) two possible {\bf dO-cO} co-existences where condensed and dilute phases are transported in the same direction (blue) or in opposite directions (red); in both cases shown here, the condensed ordered phase is moving to the right as the magnitude of the flux in the condensed phase, $|J_{\rm c}|$, is higher than that in the dilute phase $|J_{\rm d}|$; (c) two examples of {\bf dD-cO} co-existence, where in both cases, the condensed ordered phase is moving to the right; and (d) {\bf dO-cD} co-existence, the condensed disordered phase is here moving to the left due to differential adsorption at the two interfaces. The profiles shown here correspond to the stationary states of the hydrodynamic EOM. In the lower panels, the symbols correspond to the symbols shown in Fig.\,\ref{fig:phasediagram}. Movies of these cases can be found in \cite{SM}.}
\label{fig:profiles}
\end{figure*}

{\it Conservation law and symmetries.---}Our model EOM are based on the conservation law and symmetries in the system. Specifically, mass conservation leads to the continuity equation:
\beq
\pp_t \rho +\vnab \cdot {\bf J}=0\ ,
\label{eq:density}
\eeq
where the total flux ${\bf J} = \bp - \eta \nabla \rho$ 
is composed of an {\it active flux} $\bp$ and a {\it gradient term} (leading to a diffusive term in the continuity equation). 

For the EOM of the active flux $\bp$, following \cite{toner-prl-1995,toner-pre-1998,toner-annphys-2005,toner-pre-2012}, we impose temporal, translation, rotation, and chiral invariances to obtain:
\beq
\label{eq:momentum}
\pp_t \bp +\lambda \bp \cdot \nabla \bp = \mu \nabla^2 \bp - \kappa (\rho)\nabla \rho +\alpha(\rho) \bp -\beta p^2 \bp 
\ ,
\eeq
where we have only retained the terms crucial to our discussion here. We have also emphasized the density dependency of the ``compressibility" coefficient $\kappa$ and that of the ``order-disorder'' control parameter $\alpha$ in the above equations.

Note that our EOM differ from the Toner-Tu EOM in our choice of hydrodynamic variables and the imposition of the diffusive term in the EOM of $\rho$. Indeed, while $\rho$ describes our active fluid mass density, $\mathbf{p}$ denotes here an active flux and can only formally be identified with the momentum density when the diffusive term vanishes ($\eta \to 0$), in which case our EOM reduce exactly to the reduced Toner-Tu EOM. The presence of this diffusive term facilitates our numerical analyses of the EOM and was commonly adopted in previous studies \cite{fily-prl-2012,farrell-prl-2012,dunkel-njp-2013,worlitzer-arxiv-2020}. However, since we recover diverse salient features known in polar active fluids, we are confident that the findings in our paper remain valid for generic active fluids as described by the Toner-Tu EOM. In particular, we show in \cite{SM} that our results are qualitatively unchanged by the presence of the diffusive term.

{\it Diversity of phase separations.---} Phase separation occurs in systems with a conserved quantity. Here, the conserved quantity is the total mass and so a phase co-existence consists of one condensed density phase (denoted by {\bf c}) and one dilute density phase (denoted by {\bf d}). At the same time, the Toner-Tu model allows for two distinct spatially homogeneous phases: the disordered ({\bf D}) and the ordered ({\bf O}) phases, characterized by whether the non-conserved order parameter $|\la \bp \ra|$ 
is zero or not. We therefore generically expect four possible phase co-existences: (i) {\bf dD-cD} (i.e., a dilute disordered phase co-existing with a condensed disordered phase), (ii) {\bf dD-cO}, (iii) {\bf dO-cD}, and (iv) {\bf dO-cO} (see \fig \ref{fig:profiles}). Indeed, three out of these four co-existences have already been demonstrated: (i) corresponds to MIPS, (ii) corresponds to the banding regime, and (iii) corresponds to the recently uncovered reverse-banding regime  \cite{schnyder-scirep-2017,nesbitt-njp-2021,geyer-prx-2019}. To the best of our knowledge, type (iv) co-existence has never been demonstrated;  here, we first predict analytically and then confirm  its existence numerically (see Fig.\,\ref{fig:profiles}) in a particular model. To that end, we will first describe how a generic phase diagram can be constructed approximately following a linear stability analysis.

{\it Linear stability and phase separation.---}In thermal phase separation, a linear stability analysis of the dynamical equation of a phase separating system (e.g., Cahn-Hilliard equations) can reveal the spinodal decomposition region of the phase diagram of the system \cite{weber-rpp-2019}. Furthermore, a signature of phase separation is that the most unstable mode from  the linear instability analysis corresponds to the $k \rightarrow 0$ mode where $k$ is the wavenumber. We will use these criteria as our guiding principles in constructing an approximate phase diagram for a particular hydrodynamic model. Specifically, we will perform a linear stability analysis on the EOM and focus on the $k \rightarrow 0$ limit. 
Furthermore, since in the disordered phase, the instability has no direction dependency; while in the ordered phase, the most unstable direction is longitudinal to the direction of the collective motion, the initial perturbation in our stability analysis is therefore taken to be along the direction of the ordered state \cite{bertin-pre-2006,bertin-jphysa-2009}. We note that all known examples of phases and phase co-existences in polar active fluids can be qualitatively understood in quasi-1D geometries; from long-ranged collective motion in ordered homogeneous phases to the one-dimensional bands observed in phase co-existences. We therefore believe that our one-dimensional analytic treatment of this problem is sufficient to capture the nature of phase co-existences in general polar active fluids, even in higher dimensions.

As an example, in the disordered case ($\alpha<0$), we expand around the homogeneous disordered state with
$\rho = \rho_0 + \delta \rho \exp [st -  \ii kx]$, $p =  \delta p \exp[st -  \ii kx]$, where we have arbitrarily chosen the $x$ direction to be the direction of interest. We note that in the Toner-Tu theory, symmetries generically allow all the phenomenological coefficients appearing in Eq.\,(\ref{eq:momentum}) to be functionally dependent on the density $\rho$, and we now proceed to Taylor expand $\kappa$ and $\alpha$ in (\ref{eq:momentum}) as follows:
\beq
\label{eq:kappa}
\kappa(\rho) = \sum_{i=0}^\infty \kappa_i \delta \rho^i
\quad \mathrm{and} \quad
\alpha(\rho) =\sum_{i=0}^\infty \alpha_i \delta \rho^i\,.
\eeq
To linear order, the EOM read:
\beqn
\pp_t \delta \rho &=& -\pp_x \delta p + \eta \pp^2_x \delta \rho
\ ,
\\
\nonumber
\pp_t \delta p  &=& \mu \pp_x^2 \delta p -\kappa_0\pp_x \delta \rho  - |\alpha_0| \delta p
\ .
\eeqn
Solving for $s$ and focusing on the hydrodynamic limit ($k \rightarrow 0$), we have
\begin{equation}
s = \begin{cases}
- |\alpha_0| + \left( \frac{\kappa_0}{|\alpha_0|} + \mu \right)k^2+\cO(k^3) \ , \\
- \left(\frac{\kappa_0}{|\alpha_0|} + \eta \right)k^2+\cO(k^3)
\ .
\end{cases}
\end{equation}
The first eigenvalue $(-|\alpha_0|)$ corresponds to the fast relaxation when the active flux deviates from the mean field value $p_0 = 0$ in the absence of spatial variations. The second eigenvalue quantifies when the instability sets in, which happens whenever $\kappa_0 + \eta |\alpha_0|$ becomes negative. Since the system is in the disordered regime, within this instability region, the system exhibits {\bf dD-cD} coexistence as shown in Fig.\,\ref{fig:profiles}(a).

The analysis in the ordered regime ($\alpha>0$) follows  the exact same procedure; the full details of the linear stability analysis can be found in \cite{SM}. Here, we just recall that when $\kappa - \eta \alpha > 0$, the homogeneous disordered phase will generically be separated from the homogeneous ordered phase by phase co-existence regions. We will now present a particular model that illustrates the diversity of phase transitions and phase co-existences possible in active fluids.

\begin{figure}
\centering
\includegraphics[width=8.6cm]{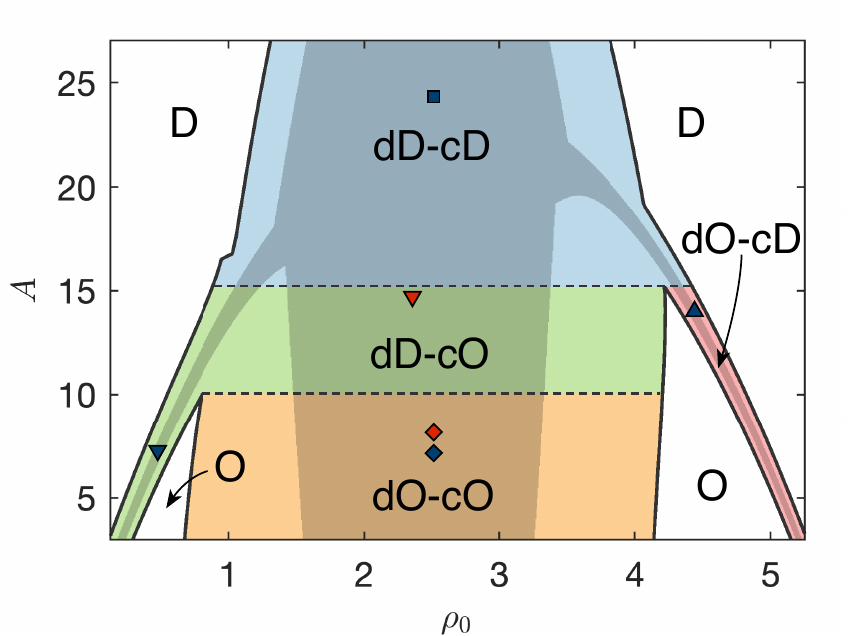}
\caption{  Phase diagram of a one-dimensional active fluid.  The shaded area corresponds to the instability regime obtained from the linear stability analysis of the hydrodynamic model (\ref{eq:alphaMIPSinCM}) and (\ref{eq:kappaMIPSinCM}). The edges of the instability region corresponds to the {\it spinodal} lines. Black lines correspond to the {\it binodal} lines (or co-existence lines), which were obtained via direct numerical simulations of the one-dimensional hydrodynamic model in (\ref{eq:alphaMIPSinCM}) and (\ref{eq:kappaMIPSinCM}). Along with the homogeneous disordered ({\bf D}) and homogeneous ordered ({\bf O}) phases, we observe four co-existence regions which are delimited by the binodal lines (and horizontal dashed lines): a co-existence of a dilute disordered phase with a condensed disordered phase ({\bf dD-cD}, blue), a co-existence of a dilute disordered phase with a condensed ordered phase ({\bf dD-cO}, green), a co-existence of a dilute ordered phase with a condensed disordered phase ({\bf dO-cD}, red) and a {\it novel} co-existence of a dilute ordered phase with a condensed ordered phase ({\bf dO-cO}, orange). The symbols (and their colors) denote the locations in phase space from which we extracted the profiles shown in Fig.\,\ref{fig:profiles}. }
\label{fig:phasediagram}
\end{figure}

{\it A model with all four phase co-existences.---}The linear stability analysis can be applied straightforwardly once $\kappa(\rho)$ and $\alpha(\rho)$ in (\ref{eq:momentum}) are explicitly defined. Here, we consider the following model :
\begin{align}
\alpha(\rho)  &= -A+18\rho-10/3\rho^2 \label{eq:alphaMIPSinCM} \\
\kappa(\rho) &= 140 -145 \rho+ 30 \rho^2 \ , \label{eq:kappaMIPSinCM}
\end{align}
with $\eta =2$, $\lambda =1$, $\beta =0.5$ and $\mu =1$. A microscopic-level (particle based) system that realizes this model could for instance be a polar active fluid with contact inhibition of alignment (e.g., as discussed in \cite{nesbitt-njp-2021}) such that its equation of state dictates that it can also phase separates within the homogeneous ordered phase.

For large enough values of $A$, $\alpha$ remains negative, and so the system remains in the disordered phase. In this case, instability occurs if $(\kappa - \eta \alpha)<0$, and we expect {\bf dD-cD} co-existence in this parameter range. On the other hand, as $A$ decreases, the range of densities $\rho$ for which the system is in the ordered phase gets wider, and, importantly, is separated from the disordered phase by two instability regions: a {\bf dD-cO} co-existence to the left and a {\bf dO-cD} co-existence to the right. Simultaneously, $(\kappa - \eta \alpha)$ remains negative around $\rho \sim 2.5$. We therefore expect an interesting interplay of distinct phase separations.

In Fig.\,\ref{fig:phasediagram}, the instability regions resulting from our linear stability analysis are shown as the shaded area, while the homogeneous disordered ({\bf D}) and ordered ({\bf O}) regions are shown in white. We equate the instability region to be within the phase separating region, but to which of the four possible types of phase co-existences?

Since the conserved quantity here is the total mass, $\rho$ can be redistributed as long as the overall density remains the same. Therefore, we can characterize the phases as follows: given any starting point on the phase diagram within the instability region (shaded area in the phase diagram), we extend a horizontal line from that point; the first homogeneous phase encountered to the right (respectively, to the left) will describe the nature of the condensed (respectively, dilute) phase ({\bf D} or {\bf O}).

Using the above construction, we see that this particular model contains all four variations of the phase co-existences (\fig \ref{fig:phasediagram}). As aforementioned,  we report for the first time the existence of a phase co-existence in which both the dilute and condensed phases are ordered. A priori, these two co-moving phases can move either in the same direction or in opposite directions. By directly solving the hydrodynamic EOM numerically \cite{SM}, this is evidenced in the stationary profiles shown in Fig.\,\ref{fig:profiles}(b). Besides this particular model, we note that further diversity of phase diagrams are rendered possible by varying the specific definitions of $\alpha(\rho)$ and $\kappa(\rho)$; we discuss other interesting cases in \cite{SM}.

{\it Instability vs.~phase separation.---}The instability region obtained from a linear stability analysis does not correspond exactly to the whole phase separation region. Indeed, as in thermal phase separation, the instability region in fact corresponds to the spinodal decomposition region, which is always flanked by the so-called nucleation and growth regions on either side \cite{barrat-book-2003,weber-rpp-2019}. This is no different here: the actual phase separation boundaries encapsulate the instability regions (Fig.\,\ref{fig:profiles}). Of course, while the phase separation boundaries (i.e., the binodal lines) for thermal systems in equilibrium  can be obtained by analyzing the free energy, e.g., by using the Maxwell tangent method, no free energy exists in our non-equilibrium systems and so the phase separation boundaries will instead be given by the appropriate boundary conditions obtained from the stable steady-state solution of the actual hydrodynamic EOM. This is exactly what we did to obtain the profiles shown in \fig \ref{fig:profiles}. Specifically, the locations of the binodal lines correspond to the density values of the stationary regions of the condensed and dilute phases (see \cite{SM} for further details). Finally, we note that both density and active flux profiles obtained numerically display a characteristic fore-rear asymmetry with a steeper fore-front (see Fig.\,\ref{fig:profiles}). This asymmetry was already observed and discussed in both simulations of microscopic Vicsek-like and active Ising spins models, and their continuum counterparts \cite{solon-prl-2013,solon-prl-2015,solon-pre-2015b}.

\begin{figure}
\centering
\includegraphics[width=8.6cm]{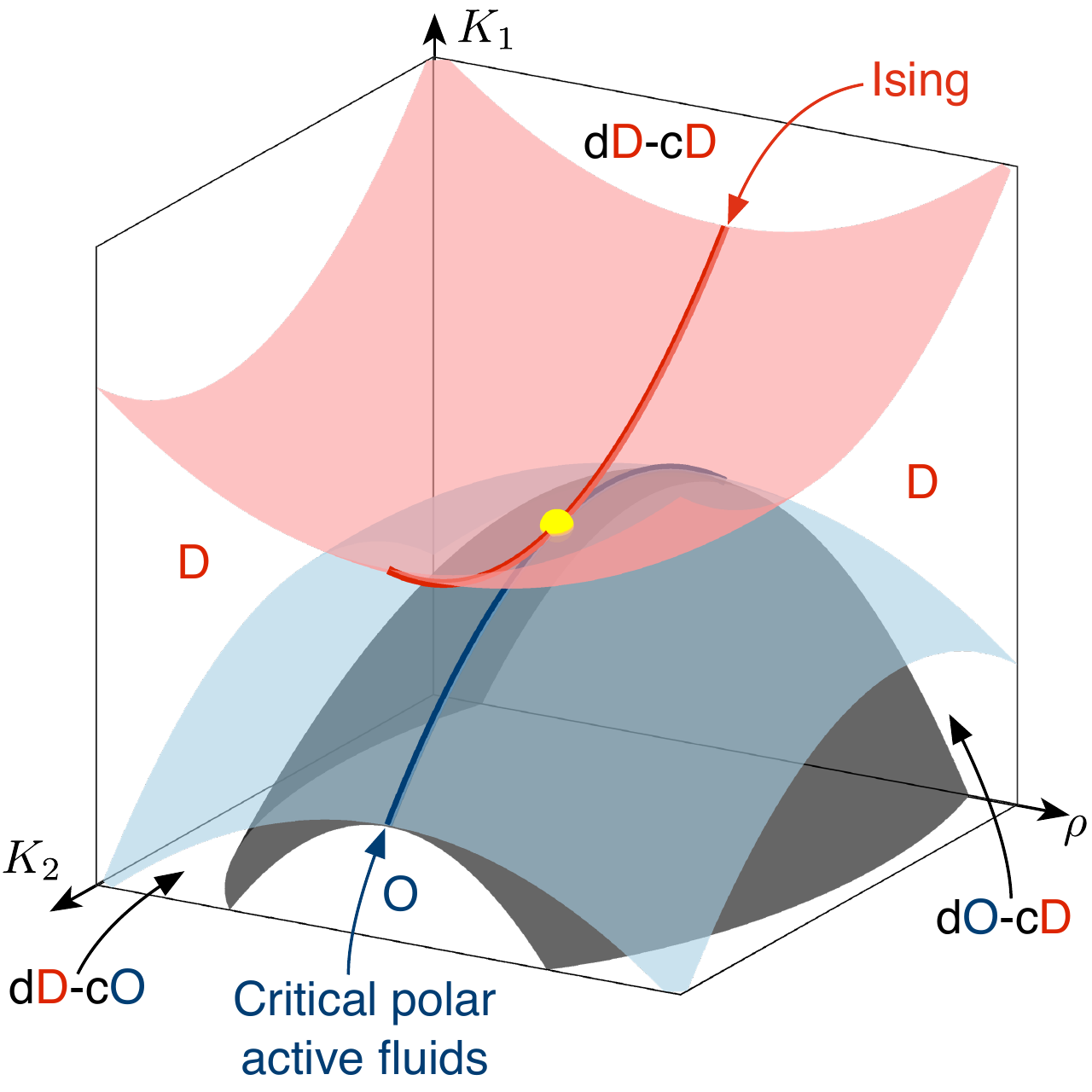}
\caption{  Multi-critical point. In this schematic, the {\bf dD-cD} co-existence (or MIPS) occurs above the red surface (at high $K_1$), and the corresponding critical phase separation (indicated by the red line) belongs to the Ising universality class \cite{partridge-prl-2019} (but see also \cite{siebert-pre-2017,caballero-jsmte-2018}). In contrast, an ordered phase co-exists with a disordered phase ({\bf dD-cO} or {\bf dO-cD}) between the blue and grey surfaces, and a spatially homogeneous ordered phase ({\bf O}) exists below the grey surface. The corresponding order-disorder critical line is indicated by the blue line \cite{nesbitt-njp-2021}. When the blue line and the red lines are tangent to each other (indicated by the yellow dot), a putatively novel multi-critical point emerges.}
\label{fig:criticalpt}
\end{figure}

{\it A multi-critical point.---}Besides uncovering the novel {\bf dO-cO} co-existence regime, our analysis also reveals a potentially novel critical point. To illustrate this (\fig \ref{fig:criticalpt}), we consider a polar active fluid system in which there are two generic parameters $K_1$ and $K_2$ that control the phase behavior of the system. Specifically, the system undergoes {\bf dD-cD} phase separation at high $K_1$ while at small $K_1$ the system is in the ordered phase. In addition, we assume that the second parameter $K_2$ controls the threshold level $K_1$ at which the distinct phase separations happen (Fig.\,\ref{fig:criticalpt}). In other words, instead of having additional phase separation due to a negative $\kappa$ inside the homogeneous ordered phase as in the previous example, we have here a {\bf dD-cD} phase co-existence in the homogeneous disordered phase instead.

Now, {\bf dD-cD} phase separation at criticality belongs to the Ising universality class \cite{partridge-prl-2019} (but see also \cite{siebert-pre-2017,caballero-jsmte-2018}). In terms of our hydrodynamic EOM,  
this critical point corresponds to having $\alpha>0, \kappa_0=\kappa_1=0$ in (\ref{eq:kappa}). On the other hand, the order-disorder critical point that accompanies critical {\bf dD-cO} and {\bf dO-cD} phase separations belongs putatively to a novel universality class ($\kappa>0, \alpha_0=\alpha_1=0$) \cite{nesbitt-njp-2021}. Therefore, by fine tuning $\kappa_0, \kappa_1, \alpha_0$ and $\alpha_1$ to zero (indicated by the yellow ball in \fig \ref{fig:criticalpt}),  these two distinct critical points coincide and the resulting multi-critical point is likely to correspond to yet a distinct universality class for the following reasons: At the linear level around this critical point, the EOM of the active flux $\bp$ is completely decoupled from that of the density field $\rho$. Specifically, the linear EOM are
\beq
\pp_t \delta \rho +\nabla \cdot \bp = \eta \nabla^2 \delta \rho
\sep
\pp_t \bp = \mu \nabla^2 \bp + \bff 
\eeq
where $\bff$ is a Gaussian noise term with a non-zero standard deviation. The fact that $\delta \rho$ does not feature in the linear EOM of $\bp$ is distinct from all  known active fluids at the order-disorder critical transition \cite{ginelli-prl-2010,peshkov-prl-2012a,chen-njp-2015,grossmann-pre-2016,mahault-prl-2018,nesbitt-njp-2021,cairoli-arxiv-2019a,cairoli-arxiv-2019b}.
 
Using the linear theory above, we can also identify some interesting novel features of this critical point. To do so, we first perform the following re-scalings:
\beqn
\bbr \mapsto \ee^{\ell} \bbr
&\sep &
t \mapsto \ee^{z\ell} t
\\
\delta \rho \mapsto \ee^{\chi_\rho \ell} \delta \rho
&\sep &
\bp \mapsto \ee^{\chi_p \ell} \bp
\ .
\eeqn 
We can then conclude that the following choice of scaling exponents leave the linear EOM invariant \cite{SM}:
\beq
z= 2
\sep
\chi_p = \frac{2-d}{2} \sep
\chi_\rho = \frac{4-d}{2}
\ .
\eeq
Applying these linear exponents to the generic nonlinear EOM then indicates that (i) the upper critical dimension $d_c$ is 6 and (ii) the first two nonlinear term that becomes relevant right below $d_c$ are $\delta \rho^2 \bp$ and $\vnab (\delta \rho^3)$ in the EOM of $\bp$  \cite{SM}.

{\it Summary \& Outlook.---}Starting from  generic hydrodynamic EOM of polar active fluids, we have unified existing phase transitions and phase separations into a single formulation. In particular, we showed that there are generically four distinct types of phase separations, and illustrated them with a particular model. In doing so, we exhibited a novel co-existence regime: the co-existence of a dilute ordered phase and a condensed ordered phase. We expect that this new phase co-existence will be observed in a microscopic model of active Brownian particles with steric repulsion and velocity-alignment interactions. The numerical study of such a microscopic model and its coarse-graining to link microscopic parameters to the phenomenological coefficients appearing in our EOM is of great interest and will be the subject of further studies. Moreover, we also revealed a putative novel critical behavior. Our work highlights the richness of generic polar active fluid models. The phase behavior can be further enriched by considering variations in other parameters in the EOM. For instance, patterns other than phase separation has been observed when the coefficient $\mu$ in the EOM of the momentum density field (\ref{eq:momentum}) becomes negative \cite{wensink-pnas-2012}. We believe that elucidating these diverse phase behaviors will be a fruitful research direction in the future.

\begin{acknowledgments}
We are grateful to Patrick Jentsch for correcting a mistake regarding the multi-critical point in an earlier version of the paper.
\end{acknowledgments}


%

\end{document}


\title{Supplemental Material for ``Diversity of phase transitions and phase separations in active fluids"}

\author{Thibault Bertrand}
\email[Electronic address: ]{t.bertrand@imperial.ac.uk}
\affiliation{Department of Mathematics, Imperial College London, South Kensington Campus, London SW7 2AZ, United Kingdom}
\author{Chiu Fan Lee}
\email[Electronic address: ]{c.lee@imperial.ac.uk}
\affiliation{Department of Bioengineering, Imperial College London, South Kensington Campus, London SW7 2AZ, United Kingdom}

\maketitle

\section{Linear Stability Analysis}

In this section, we proceed to the linear stability analysis of our hydrodynamic EOM in one dimension
\begin{align}
\pp_t \rho + \pp_x p &= \eta \pp^2_x \rho \label{eq:EOM1}\\
\pp_t p +\lambda p \pp_x p &= \mu \pp^2_x p - \kappa (\rho) \pp_x \rho +\alpha(\rho) p -\beta p^3  \label{eq:EOM2}
\end{align}

\subsection{Instability of the homogeneous disordered state}

First, we consider the stability conditions for the homogeneous disordered state, i.e. a state with zero average momentum density. We linearize Eqs.\,(\ref{eq:EOM1}-\ref{eq:EOM2}) around the homogeneous disordered state with $ \rho = \rho_0 + \delta \rho$, $p = \delta p$, $\alpha(\rho) = - |\alpha_0|$ and $\kappa(\rho) = \kappa_0$. To linear order, we obtain 
\begin{align}
\pp_t \delta \rho &= - \pp_x \delta p + \eta \pp^2_x \delta \rho \\
\pp_t \delta p &= \mu \pp^2_x \delta p - \kappa_0 \pp_x \delta \rho - |\alpha_0| \delta p
\end{align}
with the fluctuation terms written as 
\begin{align}
\delta \rho &= \delta \rho_0 \exp[s t - \ii k x] \\
\delta p &= \delta p_0 \exp[s t - \ii k x]
\end{align}
Reinjecting the fluctuation terms in the linearized equations, one obtains 
\begin{align}
s \delta \rho_0 &= \ii k \delta p_0 - \eta k^2 \delta \rho_0\\
s \delta p_0 &= -k^2 \mu \delta p_0 + \ii k \kappa_0 \delta \rho_0 - |\alpha_0| \delta p_0 
\end{align}
which we can rewrite as the following eigenvalue problem 
\begin{equation}
\mathbf{A} \delta \mathbf{u}_0 = s \delta \mathbf{u}_0~,
\end{equation}
with $\delta \mathbf{u}_0 = (\delta \rho_0,\delta p_0)^\top$ and 
\begin{equation}
\mathbf{A} = \begin{bmatrix}
-k^2 \eta &  \ii k \\
\ii k \kappa_0 & -k^2\mu - |\alpha_0|
\end{bmatrix}
\end{equation}
We know that the stability conditions are given by the sign of $\operatorname{Re}[s]$; the eigenvalues of this $2 \times 2$ matrix are given by 
\begin{equation}
s = \frac{\Tr{\mathbf{A}}}{2} \pm \frac{1}{2} \sqrt{\left( \Tr{\mathbf{A}}\right)^2 - 4 \det{\mathbf{A}}}
\end{equation}
i.e. 
\begin{equation}
s = - \frac{|\alpha_0| + k^2 (\mu+\eta) }{2} \pm \frac{1}{2} \sqrt{\left[|\alpha_0| + k^2(\mu+\eta)\right]^2 -4\left[ |\alpha_0| \eta k^2 + \kappa_0 k^2 + \eta\mu k^4\right]}
\end{equation}
We know that $s$ is real as long as $\left[|\alpha_0| + k^2(\mu+\eta)\right]^2 -4k^2 \kappa_0 > 0 \iff |\alpha_0| \ge 0$ when $k \to 0$; which is true. In the hydrodynamic limit ($k \to 0$), we can expand the square root and we obtain
\begin{align}
s &= - \frac{|\alpha_0| + k^2 (\mu+\eta) }{2} \pm \frac{1}{2}\left[ |\alpha_0|^2 - 2k^2 (\mu+\eta) |\alpha_0| + (\mu+\eta)^2 k^4 -4 |\alpha_0| \eta k^2 - 4 \kappa_0 k^2 - 4 \eta\mu k^4 \right]^{1/2} \\
	   &\approx -  \frac{|\alpha_0| + k^2 (\mu+\eta) }{2} \pm \frac{1}{2} |\alpha_0| \left[ 1 - \frac{k^2 (\mu+ \eta)}{|\alpha_0|} - \frac{2 \eta k^2}{|\alpha_0|} - \frac{2k^2 \kappa_0}{|\alpha_0|^2} \right] \\
	   &\approx -\frac{1}{2} \big[|\alpha_0| \pm | \alpha_0| \big] \pm \frac{\kappa_0}{|\alpha_0|}k^2 - \frac{1}{2}\left[(\mu+\eta) k^2 \pm (\mu+\eta) k^2 \right] \pm \eta k^2 + \cO(k^3)
\end{align}
Finally, this leads to the following two eigenvalues in the hydrodynamic limit
\begin{equation}\boxed{
s = \begin{cases}
- |\alpha_0| + \left[\frac{\kappa_0}{|\alpha_0|} + \mu \right]k^2+\cO(k^3)  \\
-\left[\frac{\kappa_0}{|\alpha_0|} + \eta \right]k^2+\cO(k^3)
\end{cases}}
\end{equation}

We interpret these solutions as follows: 
\begin{itemize}
\item the first eigenvalue $-|\alpha_0|$ characterizes the {\it fast relaxation of the momentum fluctuations in the absence of spatial variations} ($k \to 0$);
\item the second eigenvalue $-k^2 \left[\kappa_0/|\alpha_0| + \eta \right]$ controls the {\it onset of instability of the homogeneous disordered phase}. 
\end{itemize}
We conclude that the homogeneous disordered phase is unstable when $\alpha(\rho)<0$ and $\kappa(\rho)-\eta \alpha(\rho)<0$ leading to phase separation and the emergence of the cD-dD phase co-existence (e.g. MIPS).

\subsection{Stability of the homogeneous ordered phase}
\label{sec:HOstability}

The results of the previous section hint at the fact that collective motion is expected in the case the momentum fluctuations do not relax quickly, i.e. when $\alpha(\rho) > 0$. Indeed, it is interesting to note that for a stable system $\beta>0$ necessarily, thus $\alpha_0>0$ leads to collective motion, while $\alpha_0 \le 0$ gives a stationary state. Naturally, the next step is thus to explore the stability of a homogeneous state displaying collective motion, i.e. a homogeneous ordered state. The steady-state homogeneous solutions are given by 
\begin{equation}
\beta p^3 - \alpha_0 p = 0 \iff p_0 = 0 \mathrm{~or~} p_0 = \sqrt{\alpha_0/\beta}
\end{equation}

Here, we thus expand about a homogeneous state displaying collective motion with
\begin{align}
\begin{cases}
\rho = \rho_0 + \delta \rho  &= \rho_0 + \delta \rho_0 \exp [s t -  \ii kx],	\\
p =  p_0 + \delta p &= p_0 + \delta p_0 \exp[ s t -  \ii kx], \\
 \alpha  = \alpha'_0 + \alpha'_1 \rho &= \alpha_0 + \alpha_1 \delta \rho_0 \exp[s t - \ii kx].
 \end{cases}
 \end{align}
 with $p_0 = \sqrt{\alpha_0 /\beta}$. To linear order, we obtain for the momentum equation
\begin{align}
 \pp_t \delta p + \lambda p_0 \pp_x \delta p &= \mu \pp^2_x \delta p - \kappa \pp_x \delta \rho + (\alpha_0 + \alpha_1 \delta \rho)(p_0 + \delta p) - \beta (p_0 + \delta p)^3 \nonumber \\
 								  &= \mu \pp^2_x \delta p - \kappa \pp_x \delta \rho + \alpha_0 p_0 + \alpha_1 p_0 \delta \rho + \alpha_0 \delta p - \beta (p_0^3 + 3 p_0^2\delta p) \nonumber \\
								  &= \mu \pp^2_x \delta p - \kappa \pp_x \delta \rho + \alpha_0 p_0 + \alpha_1 p_0 \delta \rho + \alpha_0 \delta p - \alpha_0 p_0 - 3 \alpha_0 \delta p \nonumber \\
								  &= \mu \pp^2_x \delta p - \kappa \pp_x \delta \rho + \alpha_1 p_0 \delta \rho + \alpha_0 \delta p - 3 \alpha_0 \delta p \nonumber 
\end{align}
The linearized equations of motion thus finally read
\begin{align}
\pp_t \delta \rho &= -\pp_x \delta p + \eta \pp^2_x \delta \rho\\
 \pp_t \delta p + \lambda p_0 \pp_x \delta p &= \mu \pp^2_x \delta p - \kappa \pp_x \delta \rho + \alpha_1 p_0 \delta \rho - 2 \alpha_0 \delta p
\end{align}
 Reinjecting in these linearized equations the ansatz, we get 
 \begin{align}
 s \delta \rho_0 &= \ii k \delta p_0 - k^2 \eta \delta \rho_0 \\ 
 s \delta p_0 -\ii k \lambda p_0 \delta p_0 &= -\mu k^2 \delta p_0 + \ii k \kappa \delta \rho_0 + \alpha_1 p_0 \delta \rho_0 -2\alpha_0 \delta p_0
 \end{align}
Once again, this can be written in matrix form as the following eigenvalue problem 
\begin{equation}
\mathbf{A} \delta \mathbf{u}_0 = s \delta \mathbf{u}_0~,
\end{equation}
with $\delta \mathbf{u}_0 = (\delta \rho_0,\delta p_0)^\top$ and 
\begin{equation}
\mathbf{A} = \begin{bmatrix}
-k^2 \eta &  \ii k \\
\ii k \kappa + \alpha_1 p_0 & - 2 \alpha_0 + \ii k \lambda p_0 - \mu k^2
\end{bmatrix}
\end{equation}
We know that the stability conditions are given by the sign of $\operatorname{Re}[s]$; the eigenvalues of this $2 \times 2$ matrix are given by 
\begin{equation}
s = \frac{\Tr{\mathbf{A}}}{2} \pm \frac{1}{2} \sqrt{\left( \Tr{\mathbf{A}}\right)^2 - 4 \det{\mathbf{A}}}
\end{equation}
i.e. 
\begin{align}
s = \frac{1}{2}\big[ -2\alpha_0 &+ \ii k \lambda p_0 - k^2(\mu + \eta) \big]  \nonumber \\
&\pm \frac{1}{2} \Big[ \big(-2\alpha_0 + \ii k \lambda p_0 - k^2(\mu+\eta) \big)^2 - 4(\kappa k^2 + 2 \alpha_0 \eta k^2 + \eta \mu k^4 - \ii k \alpha_1 p_0 - \ii k^3 \lambda \eta p_0) \Big]^{1/2}
\end{align}
 
Assuming that the discriminant is positive, we can expand $s$ to lowest order in $k$ to obtain the following two eigenvalues in the hydrodynamic limit ($k \to 0$)
\begin{align}
s_+ &= -\frac{\left[8 \alpha_0^3 \eta - \alpha_1^2 p_0^2 + 4 \alpha_0^2 \kappa + 2 \alpha_0 \alpha_1 \lambda p_0^2\right]k^2}{8 \alpha_0^3} + \ii \frac{k \alpha_1 p_0}{2\alpha_0} + \cO(k^3)
\end{align}
and
\begin{align}
s_- &= -2 \alpha_0 + \frac{\left[4 \alpha_0^2 (\kappa - 2 \alpha_0 \mu) - \alpha_1^2 p_0^2 + 2 \alpha_0 \alpha_1 p_0^2 \lambda \right] k^2}{8\alpha_0^3} + \ii \frac{kg_0}{2}\left[2 \lambda - \frac{\alpha_1}{\alpha_0} \right]+ \cO(k^3)
\end{align}

Finally, at the lowest order in $k$,  the real part of $s$ is thus given by 
\begin{equation}\boxed{
\operatorname{Re}[s] = \begin{cases}
-2\alpha_0 + \cO(k^2) \\
\frac{k^2}{8\alpha_0^2} \left[\frac{\alpha_1^2}{\beta} - 2\alpha_0 \left[ 2 \kappa + 4\eta \alpha_0 + \frac{\alpha_1 \lambda}{\beta} \right] \right] + \cO(k^3)
\end{cases}}
\end{equation}

We interpret these solutions as follows: 
\begin{itemize}
\item the first eigenvalue $-2\alpha_0$ characterizes the {\it fast relaxation of the momentum fluctuations around the mean field value $p_0 = \sqrt{\alpha_0 / \beta}$ in the absence of spatial variations} ($k \to 0$);
\item the second eigenvalue controls the {\it onset of instability of the homogeneous ordered phase}.
\end{itemize}

As long as $\alpha_0 > 0$, we have fast relaxation of the momentum fluctuations around the mean field value: the phase remains ordered. Instability leading to phase separation sets in when 
\begin{equation}
\frac{\alpha_1^2}{\beta} - 2\alpha_0 \left[ 2 \kappa + 4\eta \alpha_0 + \frac{\alpha_1 \lambda}{\beta} \right] > 0
\label{eq:instabilityCondition}
\end{equation}

First, we find the roots of the associated quadratic equation: $\alpha_1^2 - 2\alpha_0\lambda \alpha_1 - 4 \alpha_0 \beta \kappa - 8\eta \alpha_0^2 \beta = 0$. The discriminant of this quadratic equation is given by 
\begin{equation}
\Delta = 4 \alpha_0^2 \lambda^2 + 16\alpha_0 \kappa \beta + 32 \eta \alpha_0 \beta
\end{equation}
The sign of the discriminant dictates the existence of real roots. As $\beta >0$, if $\kappa < -[\alpha_0 \lambda^2/4\beta + 2\eta]$, the quadratic equation does not have real solutions and the inequality is always met. This means that the {\it homogeneous state is thus always unstable}.

Conversely, if $\kappa > -[\alpha_0 \lambda^2/4\beta + 2\eta]$, the roots of the quadratic equation are given by 
\begin{equation}
\alpha_1 = \alpha_0 \lambda \pm \sqrt{\alpha_0^2 \lambda^2 + 4\alpha_0 \kappa \beta + 8 \eta \alpha_0 \beta}
\end{equation}

We conclude that if $\kappa > -[\alpha_0 \lambda^2/4\beta + 2\eta]$, the instability sets in only when 
\begin{equation}
\alpha_1 < \alpha_0 \lambda\left[ 1 - \sqrt{1 + \frac{4 (\kappa+2\eta) \beta}{\alpha_0 \lambda^2}} \right] \mathrm{\quad or \quad} \alpha_1 > \alpha_0 \lambda\left[ 1 + \sqrt{1 + \frac{4 (\kappa+2\eta) \beta}{\alpha_0 \lambda^2}} \right]
\end{equation}

Under these instability conditions, the homogeneous ordered phase is unstable. We will see in what follows that in this case, the instability can lead to the emergence of one of the three remaining phase co-existences: cO-dD ({\it banding}), cD-dO ({\it reverse banding}) or cO-dO ({\it comoving phases}). Interestingly, in the ordered phase, the homogeneous state can remain stable even if $\kappa$ is negative, as long as the above stability conditions are satisfied. 
 
\section{Particular models}

\subsection{A first example}

In general, we are free to choose any functional form for $\alpha(\rho)$ and $\kappa(\rho)$. Here, we consider a specific model and use the previous results to study when this model can lead to non-trivial phase co-existence. We consider the following model
\begin{align}
\alpha(\rho)&= -A+\rho-\rho^2 \label{eq:alpha} \\
\kappa(\rho) &= K-\frac{5}{6}\rho +\rho^2 \label{eq:kappa}
\end{align}
and $\lambda = \mu = \beta = \eta = 1$.

We have seen above that the condition for existence of collective motion is $\alpha(\rho) > 0$. In this model, $\alpha$ can become positive when the discriminant of the quadratic equation $\rho^2 - \rho + A = 0$ is positive, i.e. when $A<1/4$. We conclude that collective motion is possible in this model when
\begin{equation}
\frac{1}{2}\left(1-\sqrt{1-4 A}\right)< \rho < \frac{1}{2}\left(1+\sqrt{1-4 A}\right)
\end{equation}

In particular, within the region where $\alpha(\rho)<0$, we can linearize the expression of $\alpha$ and write
\begin{equation}
\alpha(\rho) = (-A+\rho_0-\rho_0^2) +(1-2\rho_0) \delta \rho +\cO(\delta \rho^2)
\end{equation}
where we have introduced $\rho = \rho_0+\delta \rho$. Thus, identifying this expression to the results we derived in Section \ref{sec:HOstability}, we write $\alpha(\rho) = \alpha_0 + \alpha_1 \delta \rho$ with 
\begin{align}
\alpha_0 &= (-A+\rho_0-\rho_0^2) \\
\alpha_1 &= (1-2\rho_0)
\end{align}

Substituting this into Equation (\ref{eq:instabilityCondition}), we obtain the following condition
\begin{equation}
\frac{(1-2\rho_0)^2}{\beta}-2(-A+\rho_0-\rho_0^2)\left[2 \left( K-\frac{5\rho_0}{6}+\rho_0^2\right) + 4\eta \left(-A+\rho_0-\rho_0^2 \right) + \frac{1-2\rho_0}{\beta}\right]>0
\end{equation}

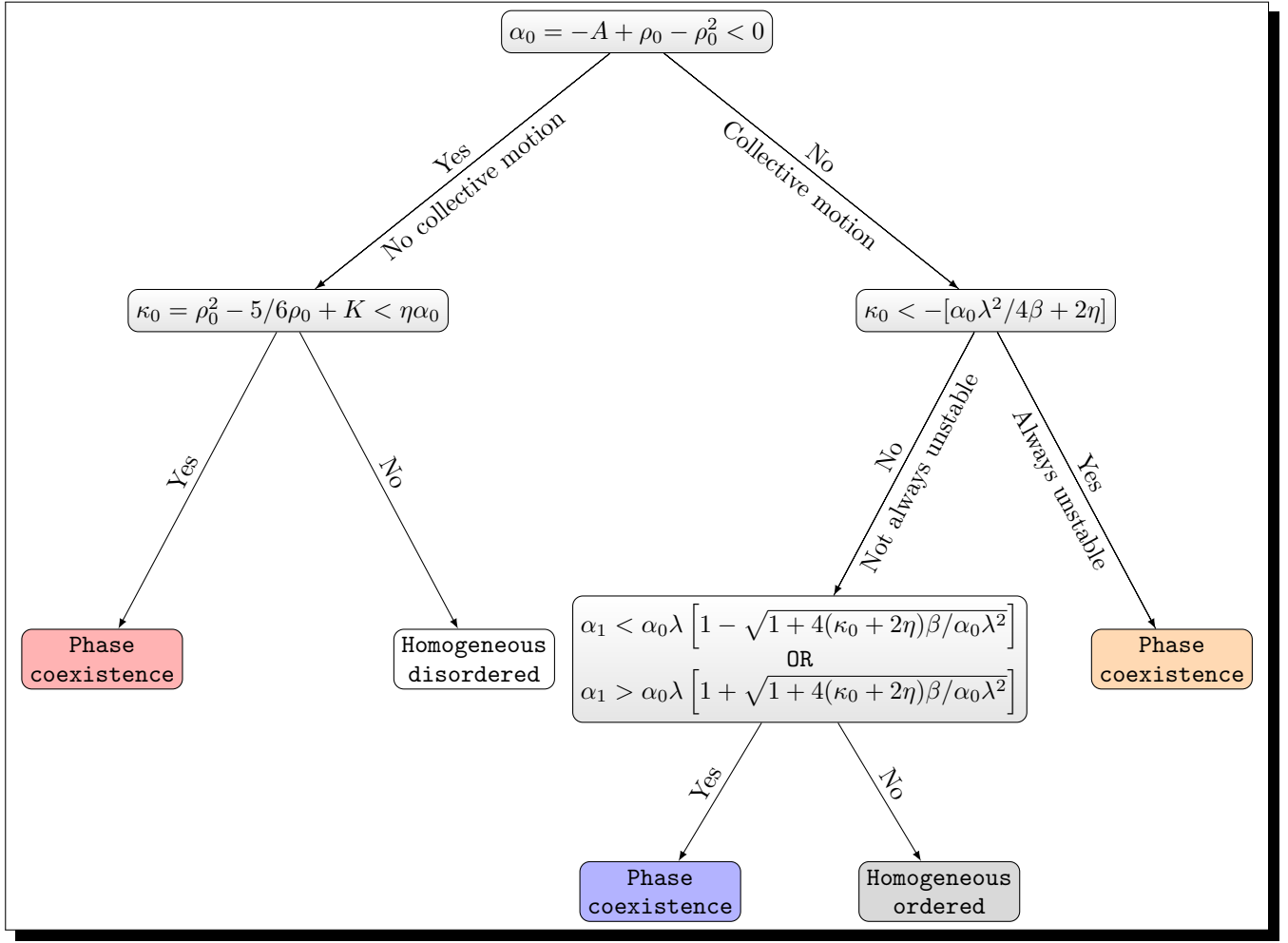
\begin{figure}[b!]
\centering
\shadowbox{
\begin{tikzpicture}
  [
    grow                    = south,
    level 1/.style={level distance=12em, sibling distance=30em},
    level 2/.style={level distance=15em, sibling distance=16em},
    level 3/.style={level distance=10em, sibling distance=12em},
    level 4/.style={level distance=12em, sibling distance=10em},
    edge from parent/.style = {draw, -latex},
    every node/.style       = {font=\normalsize},
    sloped
  ]
  \node [env] {$\alpha_0 = -A + \rho_0 - \rho_0^2 < 0$}
  	child { node [env] {$\kappa_0 = \rho_0^2 - 5/6 \rho_0 + K < \eta \alpha_0$}
		child { node [env1] {Phase \\ coexistence}
		edge from parent node [above] {Yes}}
		child { node [env0] {Homogeneous\\disordered}
		edge from parent node [above] {No}}
	edge from parent node [above] {Yes} 
	edge from parent node [below] {No collective motion}} 
	child { node [env] {$\kappa_0 < - [\alpha_0 \lambda^2/4\beta + 2\eta]$}
		child { node [env] {$\alpha_1 < \alpha_0 \lambda\left[ 1 - \sqrt{1 + 4 (\kappa_0+2\eta) \beta/\alpha_0 \lambda^2} \right]$ \\ OR \\ $\alpha_1 > \alpha_0 \lambda\left[ 1 + \sqrt{1 + 4 (\kappa_0+2\eta) \beta/\alpha_0 \lambda^2} \right]$}
			child { node [env3] {Phase \\ coexistence}
			edge from parent node [above] {Yes}}
			child { node [env4] {Homogeneous\\ordered}
			edge from parent node [above] {No}}
		edge from parent node [above] {No}
		edge from parent node [below] {Not always unstable}}		
		child { node [env2] {Phase\\coexistence}
		edge from parent node [above] {Yes}
		edge from parent node [below] {Always unstable}}
	edge from parent node [above] {No} 
	edge from parent node [below] {Collective motion}};
\end{tikzpicture}
}
\caption{Conditions for instability stemming from the linear stability analysis; the colorscheme used corresponds to the color of the instability regions in the phase diagrams.} \label{fig:condinstability}
\end{figure}

\begin{figure}
\begin{center}
\includegraphics[width=0.72\textwidth]{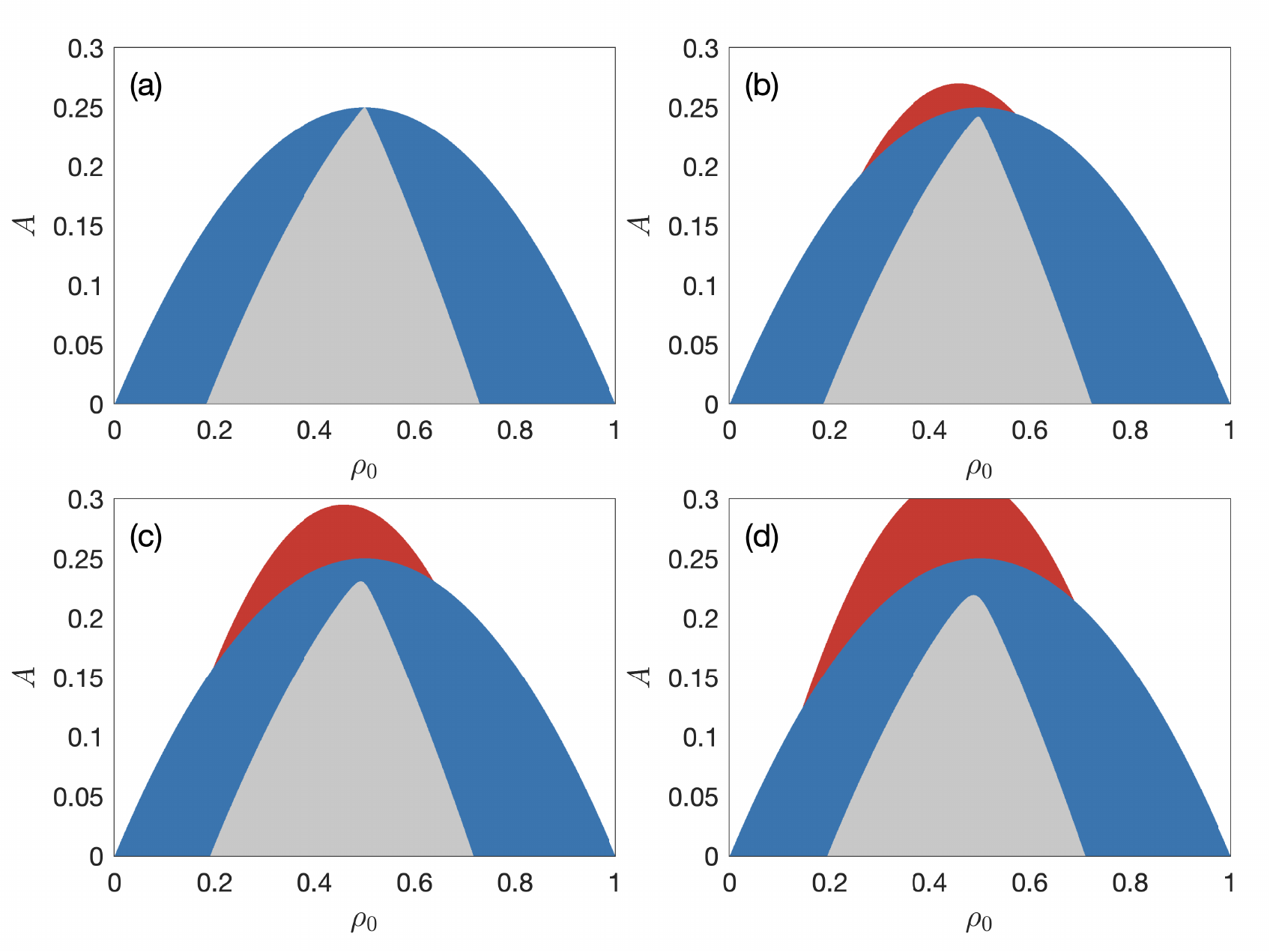}
\end{center}
\caption{Instability regions for the model defined by Eqs. (\ref{eq:alpha}) and (\ref{eq:kappa}) with fixed $K$ with values (a) $0.175$, (b) $0.15$, (c) $0.125$ and (d) $0.1$.}
\label{fig:fixedK}
\end{figure}

\begin{figure}
\begin{center}
\includegraphics[width=0.72\textwidth]{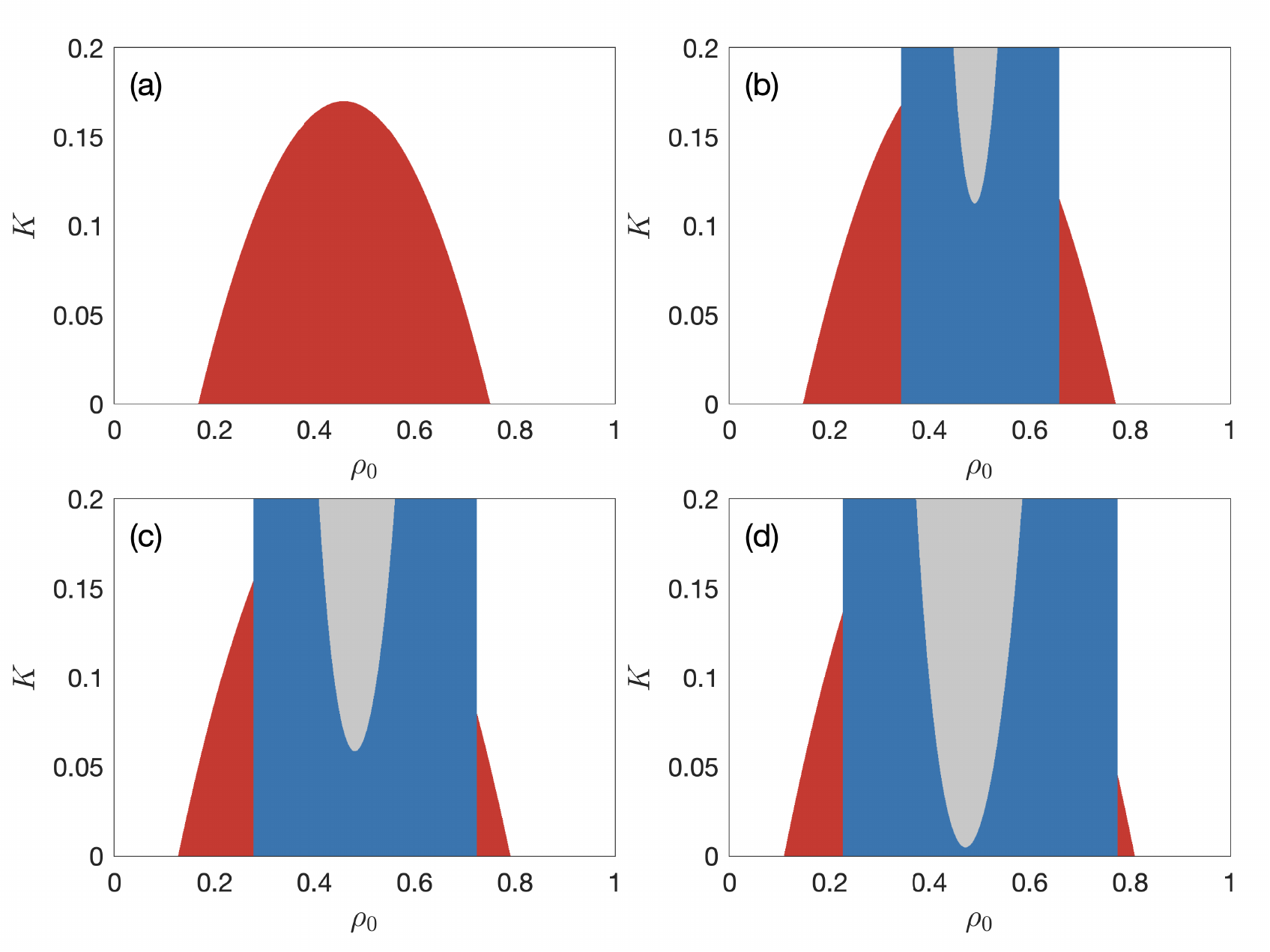}
\end{center}
\caption{Instability regions for the model defined by Eqs. (\ref{eq:alpha}) and (\ref{eq:kappa}) with fixed $A$ with values (a) $0.25$, (b) $0.225$, (c) $0.2$ and (d) $0.175$.}
\label{fig:fixedA}
\end{figure}

The locus corresponds to the roots of a quartic equations and the analytical expressions are not illuminating. However, the instability regions can be easily obtained numerically. Figure \ref{fig:condinstability} summarizes the conditions for instability. In  Figures \ref{fig:fixedK} and \ref{fig:fixedA}, colored regions correspond to the colors in Figure \ref{fig:condinstability}, i.e.:
\begin{itemize}
\item {\bf white region}, homogeneous disordered;
\item {\bf red region}, $\alpha_0<0$ and $\kappa_0-\eta \alpha_0<0$ leading to a phase co-existence;
\item {\bf orange region}, $\alpha_0>0$ and $\kappa_0<-[\alpha_0 \lambda^2/4\beta + 2\eta]$ leading to a phase co-existence;
\item {\bf blue region}, $\alpha_0>0$ and $\kappa_0>-[\alpha_0 \lambda^2/4\beta + 2\eta]$ and $\alpha_1^2/\beta - 2\alpha_0 \left[ 2\kappa_0 + 4\eta \alpha_0 + \alpha_1 \lambda/\beta \right] > 0$ leading to a phase co-existence;
\item {\bf grey region}, $\alpha_0>0$ but no instability condition met leading to a homogeneous ordered phase.
\end{itemize} 
Note that the color scheme used here refers to the parameter ranges that lead to instability, which is distinct from the color scheme used in  Fig.\,2 in the main text, which depicts the types of phase co-existence of the system.

\clearpage
\subsection{Phase diagrams}
Using the method outlined in the main text, we can construct qualitatively the corresponding phase diagrams given the instability regions. Briefly, the crucial ingredient here is that the instability regions is analogous to the spinodal decomposition region in thermal phase separation, which is always flanked by the nucleation and growth regions, which are metastable (and hence stable under linear stability analysis). Therefore, the phase separation boundaries (or the binodal lines) always extend further from the instability boundaries. 

\begin{figure}[h!]
\begin{center}
\includegraphics[width=0.72\textwidth]{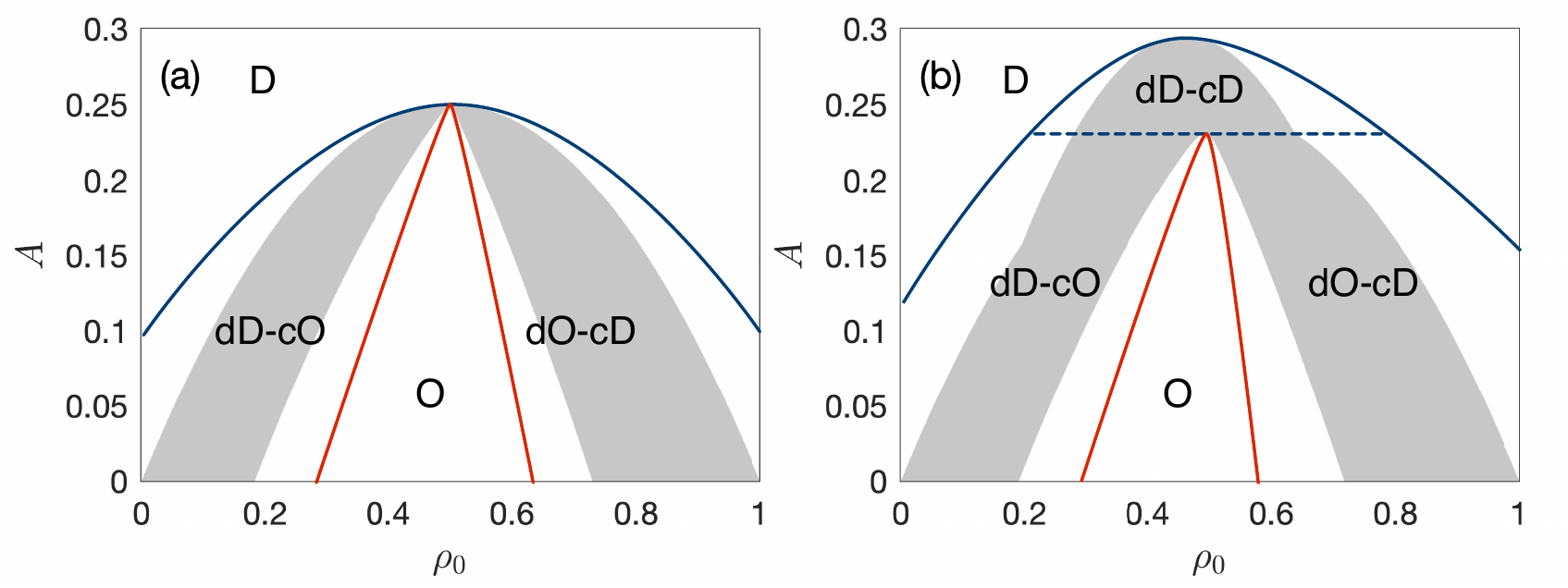}
\end{center}
\caption{Schematics of the phase diagrams of  Fig.\,\ref{fig:fixedK}(a) and (c), respectively. The different homogeneous phases and phase co-existences are separated by the blue solid, blue broken, and red lines.}
\label{fig:fixedK_PD}
\end{figure}

\begin{figure}[h!]
\begin{center}
\includegraphics[width=0.72\textwidth]{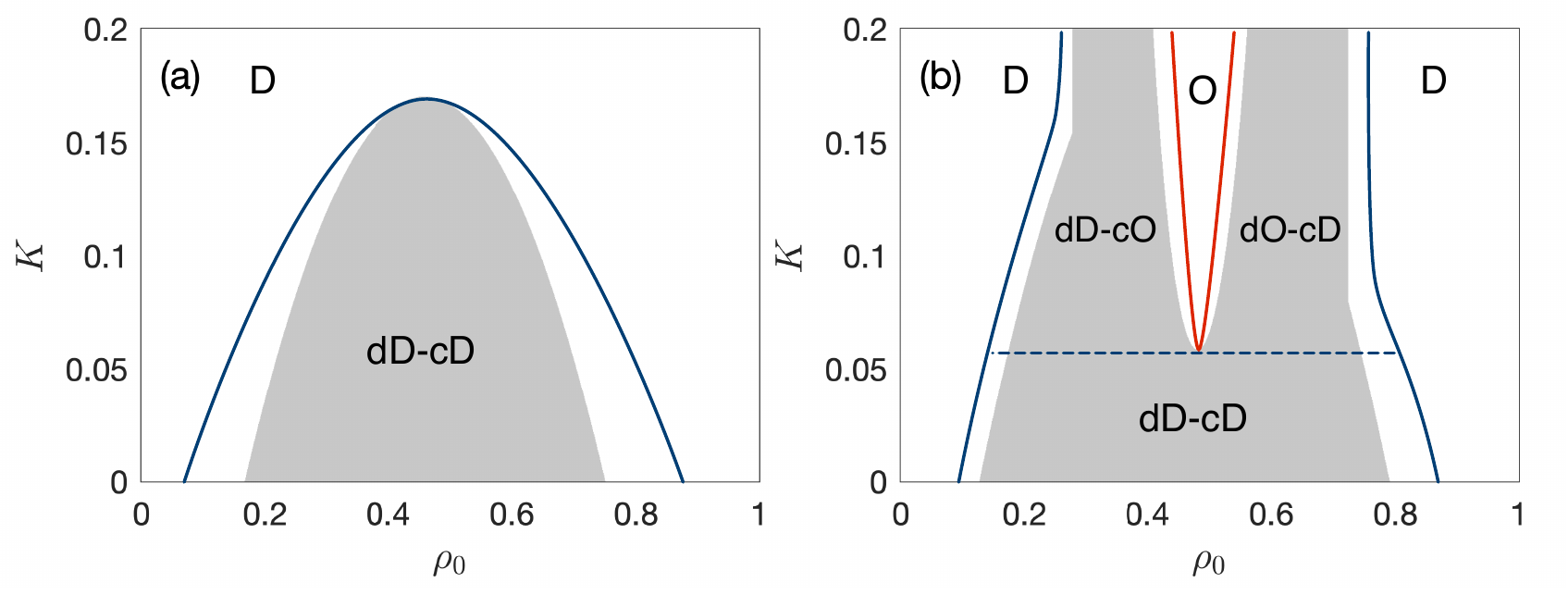}
\end{center}
\caption{Schematics of the phase diagrams of  Fig.\,\ref{fig:fixedA}(a) and (c), respectively. The different homogeneous phases and phase co-existences are separated by the blue solid, blue broken, and red lines.}
\label{fig:fixedA_PD}
\end{figure}

For instance, the corresponding phase diagrams of Fig.\,\ref{fig:fixedK}(a) and (c) are depicted qualitatively in Fig.~\ref{fig:fixedK_PD}, and those of Fig.\,\ref{fig:fixedA}(a) and (c) are shown in Fig.\,\ref{fig:fixedA_PD}. In Figs.\,\ref{fig:fixedK_PD} and \ref{fig:fixedA_PD}, all instability regions (obtained via our linear stability analysis) are shown as a shaded (grey) area while homogeneous regions are shown in white.

\subsection{A model displaying the new comoving phase co-existence}
\begin{figure}[h!]
	\begin{center}
		\includegraphics[width=0.5\textwidth]{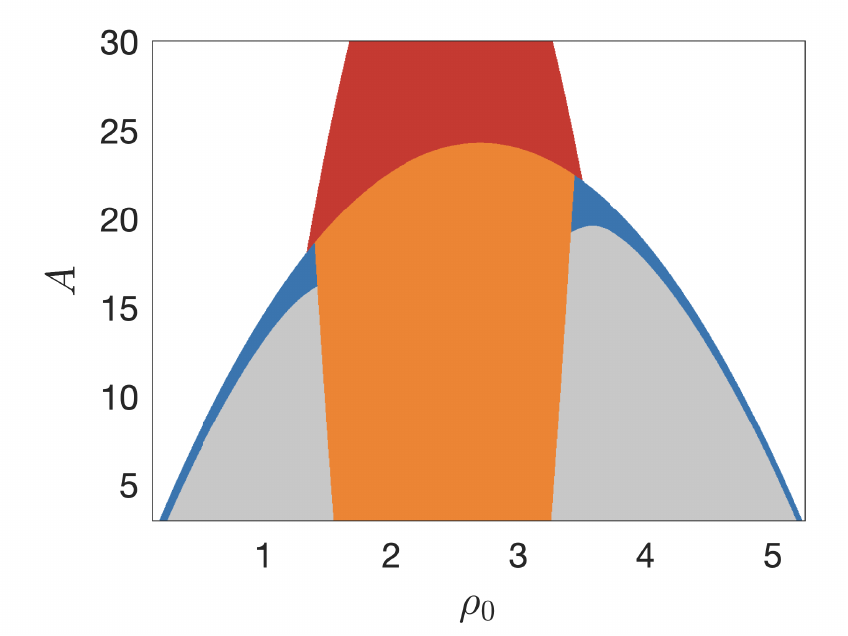}
	\end{center}
	\caption{Instability regions for the system of Equations (\ref{eq:alphaMIPSinCM}--\ref{eq:kappaMIPSinCM}) with diffusion term present in the EOM for $\rho$ ($\eta = 2$) as in the main text.} 
	\label{fig:MIPSinCM}
\end{figure}

We now discuss the model system discussed in the main text, in which
\begin{align}
\alpha(\rho)  &= -A+18\rho-10/3\rho^2 \label{eq:alphaMIPSinCM}\\
\kappa(\rho) &= 140 -145 \rho+ 30 \rho^2 
\ .\label{eq:kappaMIPSinCM}
\end{align}
The instability regions are shown in Fig.\,\ref{fig:MIPSinCM} and the corresponding phase diagram is shown in Fig.\,2 in the main text. In the next section, we provide details of the numerical methods that enables us to determine the various phase boundaries for this system.

\begin{figure}[h!]
	\begin{center}
		\includegraphics[width=0.5\textwidth]{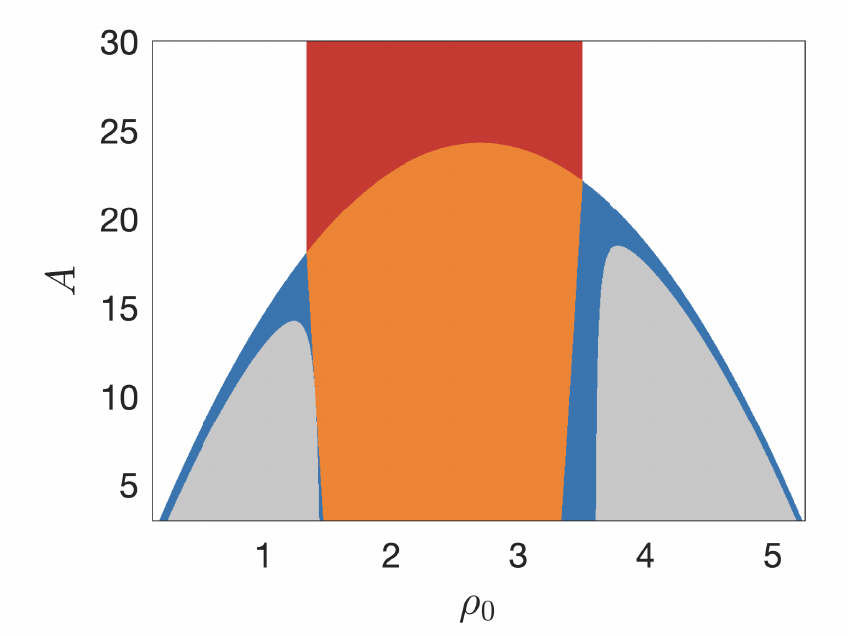}
	\end{center}
	\caption{Instability regions for the system of Equations (\ref{eq:alphaMIPSinCM}--\ref{eq:kappaMIPSinCM}) without diffusion term in the EOM for $\rho$ ($\eta = 0$).} 
	\label{fig:MIPSinCM_nodiff}
\end{figure}

Finally, we also conduct our linear stability analysis in the case where the diffusion term is not present ($\eta = 0$). We show in Fig.\,\ref{fig:MIPSinCM_nodiff} the diagram of instability regions. In particular, we find by comparing Figs.\,\ref{fig:MIPSinCM} and \ref{fig:MIPSinCM_nodiff} that the shape of the instability regions is not qualitatively impacted by the absence of the diffusion term and we thus believe that our analysis applies generally to the case of $\eta=0$ as well.

\section{Numerical Methods}

The stationary profiles displayed in the main text are obtained via direct numerical integration of Eqs.\,(\ref{eq:EOM1}) and (\ref{eq:EOM2}). To do so, we discretize space using a linearly-spaced grid with spacing $\Delta x$ and use a central finite difference approximation accurate to order $\cO(\Delta x^8)$ for the spatial derivatives. The resulting set of ordinary differential equations is integrated using a variable order implicit scheme with adaptive time-stepping (ODE15s from MATLAB --- \url{https://www.mathworks.com/help/matlab/ref/ode15s.html}). 

The results of numerical simulations shown in the main text were obtained for a domain size $L=100$ and $\Delta x \in [0.1, 0.2]$, with periodic boundary conditions. In all our simulations, we used the following parameters $\eta =2$, $\lambda =1$, $\beta =0.5$ and $\mu =1$. We made sure to reach a stable steady-state in our simulations by simulating the system for at least a total time $T=500$. To confirm the consistency of our results, we compared the steady-state profiles obtained for a variety of initial conditions (e.g., by varying the values of $\rho_c$, $\rho_d$, $p_d$, and $p_c$ as described below). For the data shown in Fig.\,1 and Fig.\,2 of the main text, we used as initial conditions a double sigmoid with sharp interfaces:
\begin{align}
\rho_1 (x) &= \rho_d + \frac{\rho_c - \rho_d}{1 + \exp \left[-(x-L/4)\right]} \nonumber \\
\rho_2 (x) &= \rho_d + \frac{\rho_c - \rho_d}{1 + \exp \left[(x-3L/4)\right]} \nonumber \\
\rho (x,0) &= \min(\rho_1(x),\rho_2(x)) \nonumber
\end{align}
and if $p_c > p_d$,
\begin{align}
p_1 (x) &= p_d + \frac{p_c - p_d}{1 + \exp \left[-(x-L/4)\right]} \nonumber \\
p_2 (x) &= p_d + \frac{p_c - p_d}{1 + \exp \left[(x-3L/4)\right]} \nonumber \\
p (x,0) &= \min(p_1(x),p_2(x)) \nonumber
\ ,
\end{align}
while if  $p_c < p_d$,
\begin{align}
p_1 (x) &= p_c - \frac{p_c - p_d}{1 + \exp \left[-(x-L/4)\right]} \nonumber \\
p_2 (x) &= p_c - \frac{p_c - p_d}{1 + \exp \left[(x-3L/4)\right]} \nonumber \\
p (x,0) &= \max\left(p_1(x),p_2(x)\right) \nonumber
\end{align}

To produce the binodal lines in Fig.\,2 of the main text, we varied $A$ (by small increments of $\Delta A = 0.25$) and initialized the system with densities taken at the edges of the corresponding instability region. We summarize in the Table \ref{tab:simparams} the parameters used for the $6$ profiles shown in Fig.\,1 of the main text and the associated supplementary movies S1 to S6.
\begin{table}[h!]
\centering
\begin{tabular}{c c c | c c | c c c c} 
 \hline
Supp. Movie & Phase co-existence & Fig. Panel & $A$ & $\rho_0$ & $\rho_d$ & $\rho_c$ & $p_d$ & $p_c$ \\ [0.5ex] 
 \hline\hline
 S1 & {\bf dD-cD} & Fig.\,1(a) & 24 & 2.5 &1.5 & 3.5 & 0 & 0 \\ 
 S2 & {\bf dO-cO} & Fig.\,1(b) (blue line) & 7 & 2.5 & 1.0 & 4.0 & 1 & 2\\
 S3 & {\bf dO-cO} & Fig.\,1(b) (red line) & 8 & 2.5 & 1.0 & 4.0 & $-1$ & 1\\
 S4 & {\bf dD-cO} & Fig.\,1(c) (blue line) & 7 & 0.45 & 0.4 & 0.5 & 0 & 1\\
 S5 & {\bf dD-cO} & Fig.\,1(c) (red line)  & 14.75 & 2.35 & 0.7 & 4.0 & 0 & 1\\ 
 S6 & {\bf dO-cD} & Fig.\,1(d) & 14 & 4.45 & 4.4 & 4.5 & 0 & $-1$\\ [1ex] 
 \hline
\end{tabular}
\caption{Simulation parameters $(A,\rho_0)$ for steady-state profiles from Fig.\,1 of the main text (and corresponding movies S1 to S6); we also include in the table the initial conditions parameters $(\rho_d,\rho_c,p_d,p_c)$.}
\label{tab:simparams}
\end{table}

\section{Multicritical Point}
At the linear level around the multi-critical point ($\kappa_0 = \kappa_1 =\alpha_0=\alpha_1=0$), the EOM are
\beq
\pp_t \delta \rho +\nabla \cdot \bp = \eta \nabla^2 \delta \rho
\sep
\pp_t \bp = \mu \nabla^2 \bp + \bff 
\ ,
\eeq
where $\bff$ is a Gaussian noise term with a non-zero standard deviation.

Performing the following re-scalings:
\beqn
\bbr \mapsto \ee^{\ell} \bbr
&\sep &
t \mapsto \ee^{z\ell} t
\\
\delta \rho \mapsto \ee^{\chi_\rho \ell} \delta \rho
&\sep &
\bp \mapsto \ee^{\chi_p \ell} \bp
\ ,
\eeqn
the EOM become
\beq
\ee^{(\chi_\rho-z)\ell} \pp_t \delta \rho +\ee^{(\chi_p-1)\ell}\nabla \cdot \bp = \ee^{(\chi_\rho-2)\ell}\eta \nabla^2 \delta \rho
\sep
\ee^{(\chi_p-z)\ell} \pp_t \bp = \ee^{(\chi_p-2)\ell} \mu \nabla^2 \bp + \ee^{-(z+d)\ell/2}\bff 
\ .
\eeq

Therefore, the above linear EOM remain invariant if we make the following choice:
\beq
z= 2
\sep
\chi_p = \frac{2-d}{2}
\sep
\chi_\rho = \frac{4-d}{2}
\ .
\eeq
Substituting these values into all possible nonlinear terms in the EOM of $\bp$ (which is identical to the Toner-Tu equation), we find that as one decreases $d$ from infinity, the first nonlinear term whose re-scaling based on the linear result diverges as $\ell \rightarrow \infty$ are  $\delta \rho^2 \bp$ and $\vnab (\delta \rho^3)$. This is because, e.g., $\delta \rho^2 \bp \mapsto \ee^{((4-d)+(2-d)/2)\ell}\delta \rho^2 \bp $, and the exponent $(5-3d/2)\ell$ becomes greater than the re-scaling exponent of  $\pp_t \bp$, which is $(\chi_\rho-z)\ell = -(d+2)\ell /2$, at the upper critical dimension $d_c=6$.